\documentclass[12pt]{iopart}
\usepackage{graphicx}
\usepackage{iopams}

\begin{document}

\title [Low temperature properties of the triangular-lattice antiferromagnet]{Low temperature properties of the triangular-lattice 
antiferromagnet: A bosonic spinon theory}

\author{A Mezio$^1$, L O Manuel$^1$, R R P Singh$^2$ and A E Trumper$^1$}

\address{$^1$ Instituto de F\'{\i}sica Rosario (CONICET) and Universidad Nacional de Rosario, Boulevard 27 de Febrero 210 bis 
(2000) Rosario, Argentina}
\address{$^2$ Department of Physics, University of California, Davis, California 95616, USA}

\ead{trumper@ifir-conicet.gov.ar}

\begin{abstract}
We study the low temperature properties of the triangular-lattice Heisenberg antiferromagnet with a mean field Schwinger 
spin-$\frac{1}{2}$ boson scheme that reproduces quantitatively the zero temperature energy spectrum derived previously using 
series expansions. By analyzing the spin-spin and the boson density-density dynamical structure factors, we identify the 
unphysical spin excitations that come from the relaxation of the local constraint on bosons. This allows us to reconstruct a 
free energy based on the physical excitations only, whose predictions for entropy and uniform susceptibility seem to be reliable 
within the temperature range $0\leq T \lesssim 0.3J$, which is difficult to access by other methods. The high values of entropy, 
also found in high temperature expansions studies, can be attributed to the roton-like narrowed dispersion at finite temperatures.
\end{abstract}

\submitto{\NJP}

\maketitle

\section{Introduction}

The study of triangular-lattice spin-$\frac{1}{2}$ Heisenberg model (THM) has been 
a central problem in quantum many body physics ever since Anderson made the
proposal that its ground state properties could be described by a Resonant Valence Bond picture \cite{Anderson73,Fazekas74}. 
The development of several numerical techniques \cite{Huse88,Bernu92,Leung93,Eltsner93,Capriotti99,Kruger06,White07,Zheng06,Mezzacapo10} 
was crucial to elucidate that the zero point quantum fluctuations, in this particular system,
are not enough to destroy the classical $120^{\circ}$ N\'eel order, lending support to a simple semiclassical description
at low temperatures, such as that provided by linear spin wave theory (LSWT)  \cite{Jolicoeur89}.

Nonetheless, early high temperature expansion (HTE) studies \cite{Eltsner93} performed down to low temperatures showed no 
evidence of a renormalized classical behaviour as predicted by the non linear sigma model (NL$\sigma$M) \cite{Azaria92,Chubukov94a}. 
For instance, around $T\!= \!0.25J$, the correlation length  calculated by HTE is only $\xi\!\sim\! 1.5$ lattice constants, in 
contrast to the value $\xi\sim12$ predicted by NL$\sigma$M \cite{Eltsner93,Zheng06}. Consistent with these values, the entropy 
calculated by HTE was one order of magnitude larger than that of the NL$\sigma$M. These early results were interpreted as a 
probable crossover between renormalized classical and quantum critical regimes \cite{Chubukov94}.

Unexpectedly, series expansion (SE) studies \cite{Zheng06,Zheng06II} performed at zero temperature also showed a strong 
downward renormalization of the high energy part of the spectrum with respect to LSWT, along with the appearance of 
roton-like minima at the midpoints of the edges of the hexagonal Brillouin zone (BZ). Originally, the presence of such 
roton-like excitations were proposed to be related to possible fermionic spinon excitations which in turn would lead to 
the anomalous low temperature properties of the THM \cite{Zheng06}. However, subsequent works showed that non trivial 
$1/S$ corrections, arising from the non collinearity of the $120^{\circ}$ N\'eel order, accurately recovers the $T\!\!=\!\!0$ 
series expansion results \cite{Starykh06,Chernyshev06,Chernyshev09}. This gave support to an interacting magnon picture for 
the spectrum, 
although it was found that the magnon-quasiparticles are not well defined for a significant part of the Brillouin 
zone \cite{Chernyshev06,Chernyshev09}. 
 Nevertheless, by assuming a bosonic character for the SE dispersion relation, the high values of entropy found at 
 low temperature was attributed to the thermal excitation of rotons, even at temperatures below the roton gap \cite{Zheng06}.   

Here we explore the low temperature properties of THM from an alternative viewpoint:
A bosonic spinon perspective, based on the Schwinger boson formalism \cite{Arovas88,Auerbach88}. One advantage of this point of view
is that it preserves the rotational invariance of the THM at finite temperatures, in agreement with the Mermin-Wagner 
theorem \cite{Mermin66}; while at zero temperature it recovers the $120^{\circ}$ N\'eel ordered state as a condensation 
of the Schwinger bosons \cite{Hirsch89,Chandra90}. In fact, it has been shown in the literature \cite{Manuel98,Gazza93} that the 
Schwinger boson theory 
describes very well the ground state properties of the THM, finding good agreement with exact 
diagonalization \cite{Lecheminant95} and other available numerical techniques (For a complete survey of the available 
results on the THM see table III of \cite{Zheng06}). More recently, we 
have shown \cite{Mezio11} that the main features of the magnetic excitation spectrum found using series expansions 
can be reproduced correctly using a proper mean field scheme of decoupling. In addition, by computing the 
density-density and spin-spin dynamical structure factors we were able to identify unphysical magnetic excitations 
that can be traced back to the relaxation of the local constraint of the Schwinger bosons at the mean field level \cite{Mezio11}. 

In practice, the relaxation of the local constraint seems to be not crucial for the correct description of certain 
ground state static properties \cite{Mezio11}. But as soon as temperature increases, the system starts to explore in 
increasingly 
large amounts an unphysical phase space, leading  to an incorrect estimation of the thermodynamic quantities. 
Instead of introducing $\it{ad\; hoc}$ factors to compensate for the above problem \cite{Arovas88,Auerbach88}, we use our ability 
to distinguish between the physical and the spurious excitations to properly extend the Schwinger boson mean field (SBMF)  to finite temperatures 
by considering only the physical low energy excitations. We rename the latter the reconstructed Schwinger boson mean 
field (RSBMF). The values of entropy and uniform susceptibility thus calculated interpolate quite well within the 
temperature range $0-0.3J$, on opposite ends of which LSWT plus $1/S$ corrections and HTE become reliable, respectively.  
Our results support the idea of \cite{Zheng06} that the high values of entropy found 
with HTE are due to the excitation of rotons which, within the context of our theory, can be identified with collinear 
short range AF fluctuations above the underlying  $120^{\circ}$ N\'eel correlations.

\section{Rotational invariant Schwinger boson mean field theory}

In using Schwinger boson representation for the spin operators \cite{Arovas88,Auerbach88},

\begin{equation}
\hat{S}^{x}_i=\frac{1}{2}(\hat{b}^{\dagger}_{i\uparrow}\hat{b}_{i\downarrow}+\hat{b}^{\dagger}_{i\downarrow}\hat{b}_{i\uparrow}),\;
\hat{S}^{y}_i=\frac{1}{2\rmi}(\hat{b}^{\dagger}_{i\uparrow}\hat{b}_{i\downarrow}-\hat{b}^{\dagger}_{i\downarrow}\hat{b}_{i\uparrow}),
\;\hat{S}^{z}_i=\frac{1}{2}(\hat{b}^{\dagger}_{i\uparrow}\hat{b}_{i\uparrow}-\hat{b}^{\dagger}_{i\downarrow}\hat{b}_{i\downarrow}),
\end{equation}
the local constraint on the boson number $\sum_{\sigma}\hat{b}^{\dagger}_{i\sigma}\hat{b}_{i\sigma}=2S$
must be imposed to fulfil the spin-$S$ algebra. The spin-spin interaction of the triangular-lattice Heisenberg model can then be 
written in terms of singlet bond operators \cite{Mezio11,Ceccatto93} as 
\begin{equation}
\hat{{\bf S}}_i \!\cdot \! \hat{{\bf S}}_j=\;: \hat{B}^{\dagger}_{ij} \hat{B}_{ij}:  -\hat{A}^{\dagger}_{ij}\hat{A}_{ij},
\label{int}
\end{equation}
where $\hat{A}^{\dagger}_{ij}\!\!=\!\!\frac{1}{2}\sum_{\sigma}\sigma \hat{b}^{\dagger}_{i \sigma}
\hat{b}^{\dagger}_{j \bar{\sigma}}$ and 
$\hat{B}^{\dagger}_{ij}\!\!=\!\!\frac{1}{2}\sum_{\sigma}\hat{b}^{\dagger}_{i\sigma}\hat{b}_{j \sigma}$ are singlet operators,  
invariant under $SU(2)$ transformations of the spinor $(\hat{b}_{i\uparrow}, \hat{b}_{i\downarrow})$ and $::$ means 
normal order. 
The identities $\hat{B}^{\dagger}_{ij}\hat{B}_{ij}\!\!=\!\!2(\hat{\bf S}_i+\hat{\bf S}_j)^2$ and 
$ \hat{A}^{\dagger}_{ij}\hat{A}_{ij}\!\!=\!\! 2(\hat{\bf S}_i-\hat{\bf S}_j)^2$ reveal the ferromagnetic and antiferromagnetic  
character of each term of (\ref{int}), respectively. Hence, the possible coexistence of both kinds of correlations 
renders this scheme of calculation ideal to investigate frustrated 
quantum antiferromagnets \cite{Manuel98,Gazza93,Mezio11,Ceccatto93,Trumper97,Flint09}.

The mean field decoupling of (\ref{int}) is performed in such way that 
$A_{ij}\!=\!\langle\hat{A}_{ij}\rangle\!=\!\langle\hat{A}^{\dagger}_{ij}\rangle$ 
and $B_{ij}\!\!=\!\!\langle\hat{B}_{ij}\rangle\!\!=\!\!\langle\hat{B}^{\dagger}_{ij}\rangle$. Then, after introducing the local constraint 
on average through  a Lagrange 
multiplier $\lambda$, the diagonalized mean field Hamiltonian becomes
\begin{equation}
\hat{H}_{\rm MF}= E_{\rm gs}+\sum_{\bf k} \omega_{\bf k} \left[\hat{ \alpha}^{\dagger}_{{\bf k}\uparrow} 
\hat{\alpha}_{{\bf k}\uparrow}+
\hat{\alpha}^{\dagger}_{-{\bf k}\downarrow} \hat{\alpha}_{-{\bf k}\downarrow} \right].
\end{equation}
Here
\begin{equation}
E_{\rm gs}=\frac{1}{2}\sum_{\bf k}\omega_{\bf k}+ \lambda N (S+\frac{1}{2})
\end{equation}
is the ground state energy and
\begin{equation}
\omega_{{\bf k}\uparrow}=\omega_{{\bf k}\downarrow}= \omega_{\bf k}=[(\gamma^{\rm B}_{\bf k}+\lambda)^2- 
(\gamma^{\rm A}_{\bf k})^2]^{\frac{1}{2}}
\end{equation}
is the spinon dispersion relation with geometrical factors 
$ \gamma^{\rm B}_{\bf k}\!=\! \frac{1}{2} J\sum_{\delta}  B_{\delta} \cos {\bf k}. \delta$ and 
$\gamma^{\rm A}_{\bf k}\!=\! \frac{1}{2} J\sum_{\delta} A_{\delta} \sin {\bf k}. \delta$, and 
the sums run over all the vectors $\delta$ connecting the first neighbors of the triangular lattice with $N$ sites.
The mean field parameters have been chosen real  and satisfy the relations $B_{\delta}\!=\!B_{-\delta}$ and 
$A_{\delta}\!=\!-A_{-\delta}$ \cite{Mezio11}. The mean field free energy is given by 
\begin{equation}
F = E_{\rm gs} + T \sum_{{\bf k} \sigma} \ln \left(1-e^{-\beta \omega_{{\bf k} \sigma}}\right),
\end{equation}
whose minimization with respect to the mean field parameters, $A_{\delta}$, $B_{\delta}$ and $\lambda$ leads to the following 
self-consistent equations:
\begin{eqnarray}
A_{\delta}&=& \frac{1}{2N}\sum_{\bf k} \frac{\gamma^{\rm A}_{\bf k}}{\omega_{\bf k}} \left( 1 + 2 \, n_{{\bf k}} \right) \sin{\bf k}. 
{\delta}, \nonumber \\
B_{\delta}&=& \frac{1}{2N}\sum_{\bf k} \frac{(\gamma^{\rm B}_{\bf k}+\lambda)}{\omega_{\bf k}} \left( 1 + 2 \, n_{{\bf k}} \right) 
\cos{\bf k}. {\delta}, \label{self} \\  
S+\frac{1}{2}&=& \frac{1}{2N}\sum_{\bf k} \frac{(\gamma^{\rm B}_{\bf k}+\lambda)}{\omega_{\bf k}} \left( 1 + 2 \, n_{{\bf k}} \right), 
\nonumber
\end{eqnarray}
with ${n}_{{\bf k}}=(e^{\beta {\omega}_{\bf k}}-1)^{-1}$ the bosonic occupation number. As previously stated, the advantage of 
the SBMF is to be able to study finite temperature rotationally invariant phase of the triangular 
AF, dictated by the Mermin-Wagner theorem \cite{Arovas88,Auerbach88}. In particular,  the numerical self consistent solutions of 
(\ref{self}) correspond to the renormalized classical regime with an exponential decay of the spin correlation functions 
characterized  by the magnetic wave vector ${\bf Q}=(\frac{4\pi}{3},0)$ of the $120^{\circ}$ N\'eel structure \cite{Yoshioka91}. 
This manifests in the gapped spinon dispersion $\omega_{\bf k}$ with minimum at ${\bf k}=\pm\frac{\bf Q }{2}.$ The finite 
temperature gap prevents the infrared divergences from appearing in the theory. At low temperatures the gap is exponentially 
small, becoming  $\omega_{\pm\frac{\bf Q }
 {2}}\sim1/N$ in the $T\rightarrow 0$ limit. Formally, one should first perform $N\rightarrow \infty$ and then take the limit 
 $T\rightarrow0$ to recover the $SU(2)$ broken symmetry state. In analogy with the Bose condensation phenomena, this procedure
requires one to treat separately the singular modes of (\ref{self}) and transform the sums into integrals. This gives a 
new set of equations with the quantum corrected local magnetization $m$ as a new self consistent parameter, while $\lambda$ is 
adjusted so as to get a gapless spinon dispersion in each iteration. 
Then, antiferromagnetism is interpreted as a quantum fluid where the condensate of the up/down bosons at $\pm \frac{\bf Q}{2}$  
and the normal fluid of bosons correspond to the spiraling magnetization $m$ and the zero point quantum fluctuations, 
respectively \cite{Chandra90}.
Alternatively, we have worked with large finite systems. This procedure is simpler because the same set (\ref{self}) 
are used for the $T\neq0$ and $T=0$ cases. Although the mean field solutions  keep their rotational invariant character, the 
local magnetization  $m$ can be  obtained by relating it to the static structure factor 
$S({\bf k})=\sum_{\bf R}  e^{\rmi {\bf k}. {\bf R}}\langle {\rm gs}|\hat{S}_0 \!\!\cdot\! \!\hat{S}_{\bf R}   
|{\rm gs}\rangle$ evaluated at ${\bf k}={\bf Q}$ and the singular modes as \cite{Hirsch89}
$$\frac{1}{2N}\frac{(\gamma^{\rm B}_{\frac{\bf Q}{2}}+\lambda)^2}{\omega^2_{\frac{\bf Q}{2}}}=S({\bf Q})= \frac {N}{2}m^2,$$
where the last line of (\ref{self}) has been used. For the triangular $S=\frac{1}{2}$ AF the local magnetization of the $120^{\circ}$ 
N\'eel ground state gives $m=0.275$ (We have checked that this procedure is equivalent to solve the new set of  equations above 
mentioned with self 
consistent parameters $A_{\delta}$, $B_{\delta}$ and $m$ and the sums transformed into integrals).
This value should be compared with the quantum Monte Carlo result \cite{Capriotti99}, $m_{\rm QMC}=0.205(1)$. The underestimate 
of the zero point quantum fluctuations can be attributed  to both the mean field approximation and the relaxation of the local 
constraint which violates the physical Hilbert space. In the next section, we will show that the latter leads to some unphysical 
magnetic excitations in the spectrum, requiring a careful analysis of the low lying energy excitations to properly compute  the 
low temperature properties of the THM.

\section{SBMF low lying magnetic excitations }

To analyze the magnetic excitations of the THM within the SBMF approximation we compute the  $T=0$ spin-spin  dynamic 
structure factor,      
\begin{equation}
S^{z z}({\bf k},\omega)= \sum_{n}  |\langle{\rm gs}|{\bf S}^{z}_{\bf k}|n\rangle|^2 \delta (\omega-(\epsilon_n-E_{\rm gs})),
\label{spindyn}
\end{equation}
where $|n\rangle$ are mean field excited states, and ${\bf S}^{z}_{\bf k}$ is the Fourier transforms of ${\bf S}^{z}_i$. 
We have recently shown that (\ref{spindyn}) takes the simple form \cite{Mezio11}  
\begin{equation}
S^{zz}({\bf k},\omega)\!=\!\sum_{{\bf q}} |u_{{\bf k}+{\bf q}} v_{\bf q} - u_{{\bf q}} v_{{\bf q}+{\bf k}}|^2 
\delta (\omega-(\omega_{-{\bf q}}+\omega_{{\bf k}+{\bf q}})),
\label{Skw}
\end{equation}

\begin{figure}[t]
\begin{center}
\includegraphics*[width=0.7\textwidth,angle=-0]{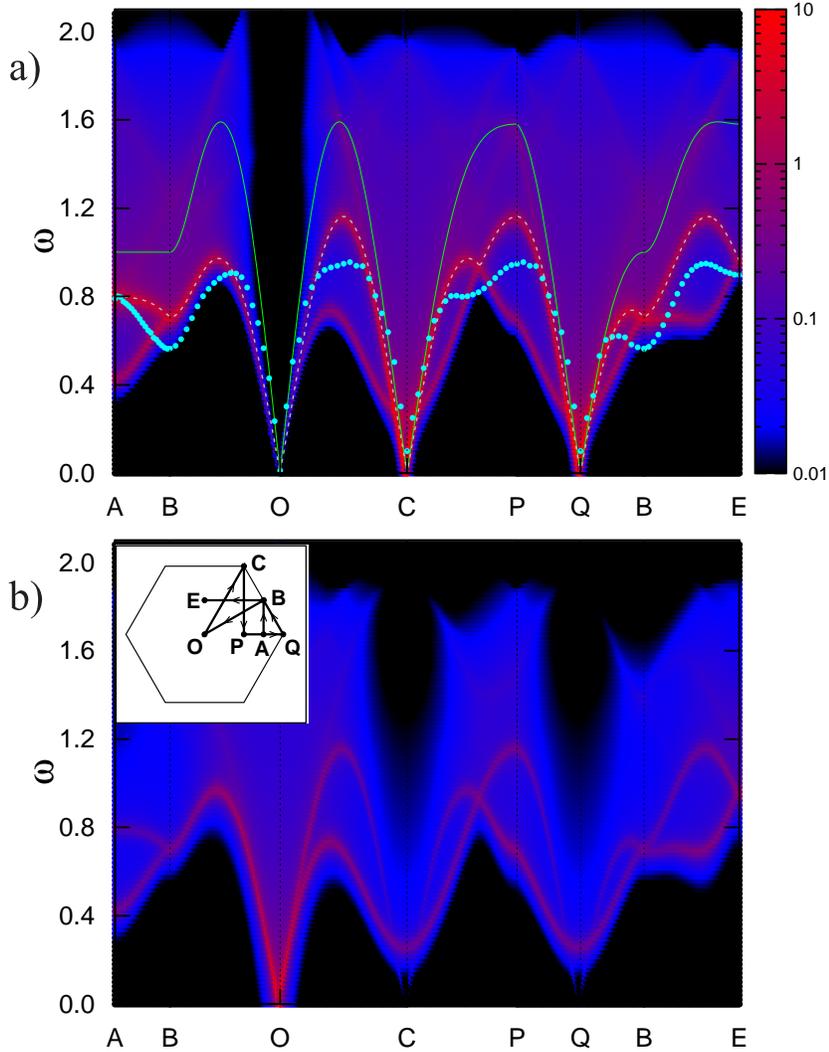}
\caption{Dynamical structure factors along the path shown in the inset of (b). (a) Magnetic structure 
factor within the Schwinger boson mean field (\ref{Skw}) (Intensity curves); LSWT results (solid green line); series 
expansion results \cite{Zheng06} (blue dots). The reconstructed dispersion relation $\overline{\omega}_{\bf k}$ is shown in dashed line. 
(b) Density-density dynamical structure factor (\ref{Nkw}). The intensity scale is logarithmic as in (a) }
\label{fig1}
\end{center}
\end{figure}

\noindent where $u_{\bf k}= [\frac{1}{2}(1+\frac{\gamma^{\rm B}_{\bf k}+\lambda}{\omega_{\bf k}} )]^{\frac{1}{2}}$ and  
$v_{\bf k}\!=\! \rmi\ \!{\rm sgn}(\gamma^{\rm A}_{\bf k})[\frac{1}{2}(-1+\frac{\gamma^{\rm B}_{\bf k}+\lambda}
{\omega_{\bf k}} )]^{\frac{1}{2}}$ are the coefficients of the Bogoliubov transformation that diagonalizes $\hat{H}_{\rm MF}$.
It is clear from (\ref{Skw}) that the spin-spin dynamical structure factor consists of two spinon
excitations which gives a continuum of spin-$1$ excitations. Furthermore, since we are working with finite systems whose mean 
field solution does not break the $SU(2)$ symmetry, $S^{x x}\!\!=\!S^{y y}\!\!=\!S^{z z}.$ Nevertheless, as the system size increases, and 
the $120^{\circ}$ N\'eel correlations are developed, there is a distinction among the two-spinon processes that can be observed 
in figure \ref{fig1}(a), where an intensity plot of (\ref{Skw}) in energy-momentum space is shown. Notice that, 
in order to show the contribution at all energies, the intensity is  plotted in a logarithmic scale.
The main signal of $S^{zz}({\bf k},\omega)$ (intense red curves) corresponds to the microscopic processes of destroying one 
spinon $b_{\pm \frac{\bf Q}{2}\sigma}$ of the condensate and creating another one $b^{\dagger}_{{\bf k}\pm\frac{\bf Q}{2} \sigma}$ 
in the normal fluid and viceversa; while the blue region corresponds to the creation of two spinons in the normal fluid 
only \cite{Mezio11}. The energy cost of the former two spin-$\frac{1}{2}$ spinon excitation is just 
$\omega_{{\bf k}\mp\frac{\bf Q}{2}}$, since in the thermodynamic limit $\omega_{\mp\frac{\bf Q}{2}}\rightarrow 0$. Notice that at 
low energies these shifted spinon based  dispersions reproduce quite well the LSWT results (solid green line) which describe the 
semiclassical long range transverse distortions of the $120^{\circ}$ N\'eel order. On the other hand, at higher energies, where the 
LSWT semiclassical description is no longer valid, the spectral weight between both shifted spinonic bands is redistributed in  
such a way that if one reconstructs a new dispersion $\overline{\omega}_{\bf k}$ from those pieces of spinon dispersions with 
dominant spectral weight (dashed line), the main features of the series expansion results \cite{Zheng06} (blue points) are recovered. There 
is, however, an additional remnant weak signal which we attribute to the relaxation of the local constraint of the mean field approximation. To quantify the relative spectral 
weight between $\overline{\omega}_{\bf k}$ and the weak band, note that in the best case --at ${\bf k}=Q$-- the contribution of 
$\overline{\omega}_{\bf k}$ represents  
around $95\%$ of the total weight while the weak band only $2\%$. On  the other hand, in the worst case -- middle point of $OC$ --  
the contributions are  $50\%$ and $20\%$, respectively.      
To describe the physical Hilbert space of the spin operators the local constraint of Schwinger bosons must be satisfied exactly, 
$\hat{{\bf S}}^2_i=\frac{n_i}{2}(\frac{n_i}{2}+1).$ So, no fluctuations of the number of boson per site should be allowed.  However, 
as the constraint is taken into account on average there appear unphysical spin fluctuations in $S({\bf k},\omega)$ 
coming from such density fluctuations which can be investigated through the density-density  dynamical structure factor defined as,
\begin{equation}
\emph{N}({\bf k},\omega)=\sum_{n}  |\langle\rm{gs}|n_{\bf k}|n\rangle|^2 \delta (\omega-(\epsilon_n-E_{\rm gs})),
\end{equation}
where $n_{\bf k}$ is the Fourier transform of $n_i=\sum_{\sigma}b^{\dagger}_{i\sigma}b_{i\sigma}$. After a little of algebra 
it results in
\begin{equation}
\emph{N}({\bf k},\omega)=\sum_{{\bf q}} |u_{{\bf k}+{\bf q}} v_{\bf q} +u_{{\bf q}} v_{{\bf q}+{\bf k}}|^2 
\delta (\omega-(\omega_{-{\bf q}}+\omega_{{\bf k}+{\bf q}})).
\label{Nkw}
\end{equation}
In figure \ref{fig1}(b), (\ref{Nkw}) is plotted in energy-momentum space. It is observed that the dominant 
spectral weight of the density-density dynamical structure factor $\emph{N}({\bf k},\omega)$ coincides with the weak signal 
of $S^{zz}({\bf k},\omega)$ (figure \ref{fig1}(a)). This notable resemblance led us to identify the remnant signal 
of $S^{zz}({\bf k},\omega)$ with the unphysical spin fluctuations originating from the density fluctuations of the Schwinger 
bosons \cite{Mezio11}.
Therefore, we expect that after projecting the mean field ground state $|{\rm gs}\rangle$ into the constrained Hilbert 
space of $2S$ bosons per site such unphysical excitations will disappear.

So far the reconstructed dispersion corresponds to a spin-$1$ excitation made of two spin-$\frac{1}{2}$ free spinons. Then, 
the relevant question is whether the spinons are bound or not once one goes beyond mean field theory. This issue has been 
addressed for frustrated AF  within the context of effective gauge field theories \cite{Read91,Sachdev91}. In particular, for 
a commensurate spinon condensed phase, it is expected that the fluctuations of the emergent gauge fields confine the spinon 
excitations, giving rise to the spin-$1$ magnon of the $120^{\circ}$  N\'eel order. While this kind of calculation is out of 
the scope of our present work, we can get some indication of the above physical picture by a simple calculation. Let us split 
the original Hamiltonian as $H\!\!=\!\!H_{\rm MF}+V$, with the interaction term given by $V\!\!=\!\!H-H_{\rm MF}$. The effect 
of $V$ on a two free spinon state $ |2{\rm s}\rangle\!\!=\!\!|{\bf q}\sigma; 
{\bf p} \sigma\rangle \!\!=\!\!\alpha^{\dagger}_{{\bf q} \sigma} \alpha^{\dagger}_{
{\bf p}\sigma}|{\rm gs}\rangle$ can be estimated, to first order in perturbation theory, by computing 
the energy of creating two spinons above the ground state as $E_{2{\rm s}}\!\!=\!\!\langle2 {\rm s}|H|2 {\rm s}\rangle\!\!-\!\!\langle{\rm gs}|H|{\rm gs}\rangle$. Then,
the interaction between the two spinons is obtained as $v_{\rm{int}}\!\!=\!\! \overline{E}_{2{\rm s}}-E^{MF}_{2{\rm s}}$ where
$E^{MF}_{2{\rm s}}\!\!=\!\!\langle2 {\rm s}|H_{MF}|2 {\rm s}\rangle-\langle{\rm gs}|H_{MF}|{\rm gs}\rangle\!\!=\omega_{{\bf q}\sigma}+\omega_{{\bf p}\sigma}$ 
and $E_{2{\rm s}}$ is rescaled as $\overline{E}_{2{\rm s}}=\frac{2}{3}E_{2{\rm s}}$ in order to compensate the difference, $\langle {\rm gs}|H|{\rm gs}\rangle\!\!=\!\!\frac{3}{2} \langle{\rm gs}|H_{MF}|{\rm gs}\rangle$, resulting 
between our mean field decoupling and the application of Wick's theorem which corresponds to a fully self consistent  Hartree-Fock-Bogoliubov 
decoupling \cite{Trumper97} (Note that this $\frac{2}{3}$ has nothing to do with the ${\it ad\;hoc}$ factor of Arovas and Auerbach 
\cite{Arovas88,Auerbach88}).
\noindent The interaction thus calculated  give

\begin{samepage}
\begin{eqnarray}
\label{interac}
 v_{\rm{int}}&=&\!\!\frac{1}{3N} \Big[\gamma_{{\bf q}-{\bf p}}\Big( u_{\bf q}^2 u_{\bf p}^2 + 
|v_{\bf q}|^2 |v_{\bf p}|^2 + 2 u_{\bf q} v_{\bf q} u_{\bf p}  v_{\bf p} \Big) + \\
\nonumber & + & 2  \gamma_{{\bf q}+{\bf p}}\Big( u_{\bf q}^2 | v_{\bf p}|^2 + u_{\bf p}^2 | v_{\bf q}|^2 - 
2 u_{\bf q}  v_{\bf q} u_{\bf p}  v_{\bf p} \Big) + 9J\Big],
\label{int}
\end{eqnarray}by the present mean field
\end{samepage}
\noindent where $\gamma_{\bf k}=  
 \frac{1}{2}\sum_{\delta} J_{\delta} \cos{\bf k}\cdot\delta$.
\noindent For a physical excitation $ |2 {\rm s}\rangle\!\!=\!\!|\pm\frac{\bf Q}{2}\sigma; {\bf k}\mp\frac{\bf Q}{2} 
\sigma\rangle$, involving the creation of one spinon in the condensate and another one in the normal fluid, it is easy to 
check numerically the attractive character of (\ref{interac}). In particular, when the total two spinon momentum is { \bf ${\bf k}\!=\!\!B$ }
 the energy binding is $v_{\rm{int}}\!\sim -0.16J$, while the energy cost to create the 
 two free spinons above the ground state is $\omega_{\frac{\bf Q}{2} \sigma}\!\!+\!\omega_{B-\frac{\bf Q}{2} \sigma}\! \sim0.7J$. 
 On the other hand, when the two spinons are created in the condensate the interaction 
 becomes $|v_{\rm{int}}|\sim O(Nm^2)$, meaning an infinite attraction for the spinons that 
 build up  the magnons at the Goldstone modes ${\bf k}=0, \pm {\bf Q}$. Even if this instability is an artifact of the first 
 order correction, we believe that this simple calculation lends support to the physical picture of tightly bound spinons in the 
 neighborhood of the Goldstone modes while at higher energies they remain weakly bound. Of course, this statement should be 
 addressed by a more rigorous calculation \cite{Chubukov91,Chubukov95}. 

\section{Low temperature properties}

As pointed out in the introduction, the SBMF ground state properties like energy, magnetic wave vector, magnetization and 
spin stiffness are not significantly affected by the relaxation of the local constraint \cite{Manuel98,Mezio11,Trumper97}. 
However, as soon as temperature increases the system starts to explore in increasingly large amounts an unphysical phase 
space due to the fluctuations of the density of Schwinger bosons. Consequently, the SBMF becomes inadequate to describe the 
low temperature properties. 
In this section we will show that it is possible to modify the SBMF in order to get reliable results for the low temperature 
regime of the THM. First, we assume that at the temperatures considered, the spinons are sufficiently bound in such way 
that the relevant physical excitations can be envisaged as spin-$1$ excitations with the reconstructed dispersion relation 
$\overline{\omega}_{\bf k}$ defined above. On the other hand, in order to mimic the projection to the physical phase space -not considered in 
the mean field- we discard the unphysical excitations. 
We have renamed this calculational scheme as reconstructed Schwinger boson mean field (RSBMF). 
We will mainly focus on the importance of the roton-like excitations to explain the high values of entropy found  
at intermediate temperatures ($T\!\sim\!0.3J$) \cite{Eltsner93,Zheng06}.
Although the RSBMF solutions correspond to the renormalized classical regime, it is worth stressing that we do not 
expect to fully recover the expected behaviour near the zero temperature transition. In fact, it is well known  that 
the correlation length at the mean field level  has an extra exponent in temperature dependence of the preexponential 
factor when compared with  the predictions of effective field theories based on the NL$\sigma$M or confined spinons near 
the zero temperature transition \cite{Chubukov91,Chubukov95}.  We have found non trivial solutions, 
$A_{\delta}, B_{\delta}\neq 0$, for the self consistent equations (\ref{self}) up to $T\sim 0.42 J$. For $T>0.42J$ the 
finite temperature phase of the triangular AF becomes a perfect paramagnet with no intersite correlations. Such a phase has no analog
in the interacting spin problem, so it is an artifact of the mean field or large-${\rm N}$ approximation. In fact, the effect of finite ${\rm N}$ corrections 
to this unwanted transition has been investigated in the literature \cite{Tchernyshyov02}. In any case, since the 
available results of HTE are reliable down to $T\sim0.3 J$ we will concentrate on the temperature range $0-0.42 J$ in 
order to interpolate between other results.   
   
\begin{figure}[t]
\begin{center}
\vspace*{0.cm}
\includegraphics*[width=0.55\textwidth,angle=-90]{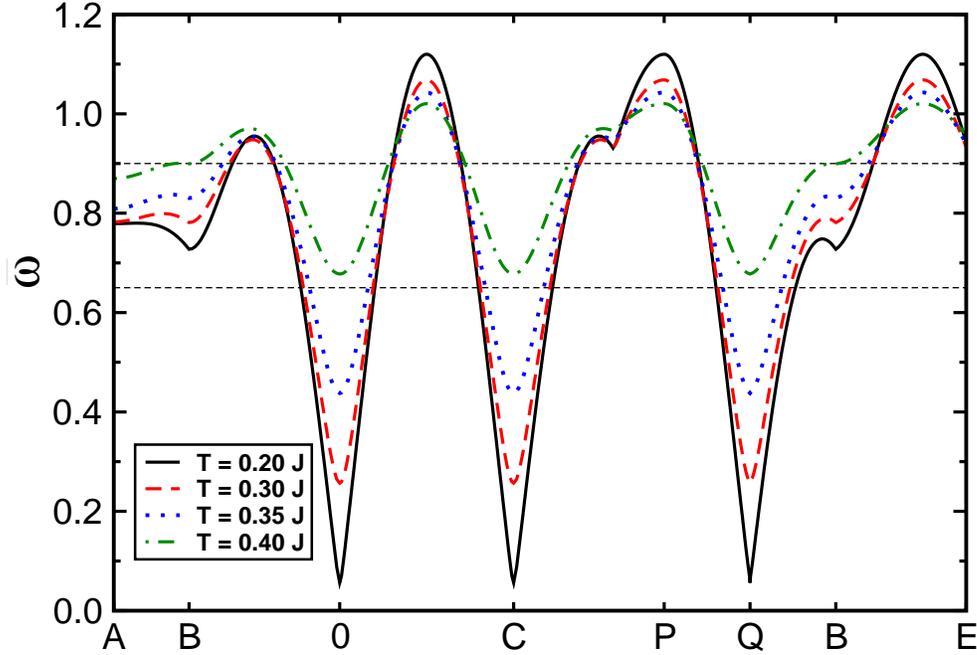}
\caption{Temperature dependence of the reconstructed dispersion relation, as defined in the text, along 
the path of the inset of figure \ref{fig1}. The horizontal dashed lines correspond to the energy intervals used in 
figure \ref{fig4}.} 
\label{fig2}
\end{center}
\end{figure}

In figure \ref{fig2} is shown the temperature dependence of the reconstructed dispersion $\overline{\omega}_{\bf k}$. As 
temperature increases, a narrowing of the bandwidth along with the opening of a gap and a flattening of the roton-like 
excitations is observed. 
\noindent The entropy per site corresponding to these low energy spin-$1$ excitations is 
\begin{equation}
\mathcal{S}  = \frac{1}{N} \sum_{{\bf k}} \left[ \; \left(\overline{n}_{{\bf k}} + 1\right) \, \ln\left(\overline{n}_{{\bf k}} + 
1\right) - \overline{n}_{{\bf k}} \; \ln \overline{n}_{{\bf k}} \; \right],
\label{SRSBMF}
\end{equation}
\noindent where the reconstructed dispersion $\overline{\omega}_{\bf k}$ is plugged in the averaged occupation number 
$\overline{n}_{{\bf k}}=(e^{\beta \overline{\omega}_{\bf k}}-1)^{-1}$. In contrast, when the two free spinon species are 
considered (SBMF) the resulting entropy per site is 
\begin{equation}
\mathcal{S}  = \frac{1}{N}  \sum_{{\bf k}\sigma} \left[ \; \left(n_{{\bf k}\sigma} + 1\right) \, 
\ln\left(n_{{\bf k}\sigma} + 1\right) - n_{{\bf k}\sigma} \; \ln n_{{\bf k}\sigma} \; \right] ,
\label{SSBMF}
\end{equation}

\begin{figure}[t]
\begin{center}
\vspace*{0.cm}
\includegraphics*[width=0.6\textwidth,angle=-90]{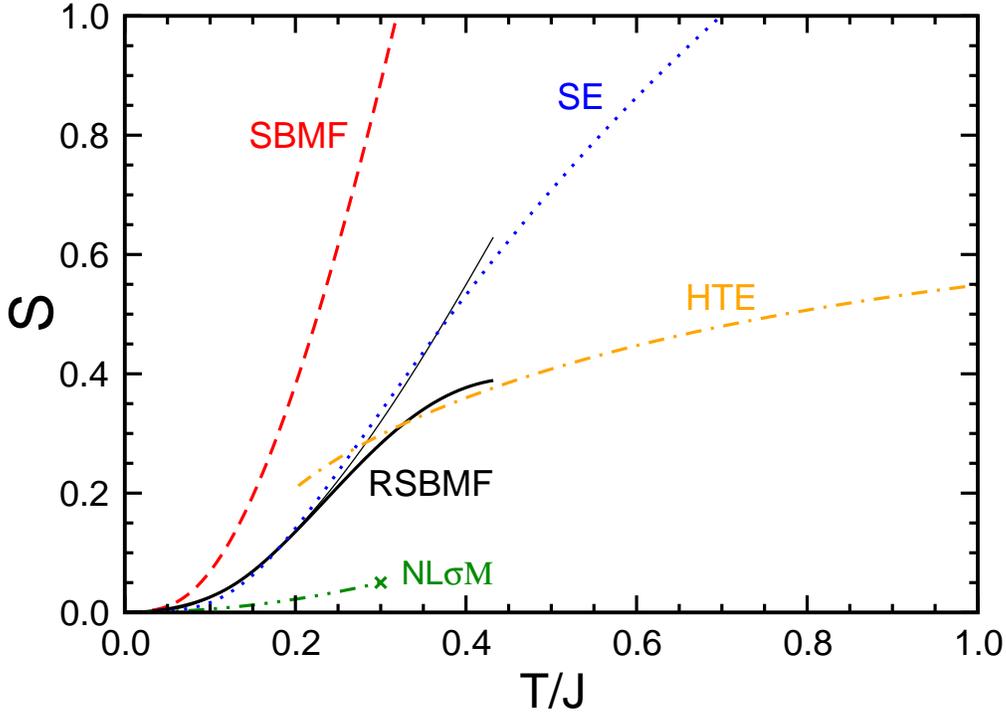}
\caption{Temperature dependence of the entropy computed within the RSBMF (\ref{SRSBMF}) (black solid line); 
SBMF (\ref{SSBMF}) (red dashed line); HTE \cite{Eltsner93} (orange dotted-dashed line); SE \cite{Zheng06}
(blue dotted line) and NL$\sigma$M  \cite{Zheng06}(green double dotted-dashed line). The black thin solid line represents 
the SBMF result divided by $2.785$ (see text in section \ref{Discussion}).}
\label{fig3}
\end{center}
\end{figure}

\noindent where  the spinon excitations $\omega_{{\bf k}\sigma}$ are plugged in the occupation number $n_{{\bf k}\sigma}$ and 
the sum over the two spinon species is taken into account. 
In figure \ref{fig3} it is shown that the RSBMF entropy (black solid line) interpolates quite well between the expected values 
at zero temperature and $T\sim0.3J$ predicted by HTE (orange dotted-dashed line); while the high values of the SBMF entropy 
(red dashed line) are due to the inclusion of the spurious excitations. To complement these results, also shown in figure 
\ref{fig3} is the entropy corresponding to the low energy spin-$1$ excitations with the dispersion relation found with SE 
at zero temperature (blue dotted line) and that of the NL$\sigma$M \cite{Zheng06}. The agreement with SE is very good at 
low temperatures, although the RSBMF entropy is better aligned with HTE at higher temperatures. This difference can be 
attributed to the fact that the SE entropy has been computed with the $T=0$ dispersion relation while in the RSBMF entropy 
the reconstructed dispersion relation is temperature dependent (see figure \ref{fig2}). 

\begin{figure}[h]
\begin{center}
\vspace*{0.cm}
\includegraphics*[width=0.6\textwidth,angle=-90]{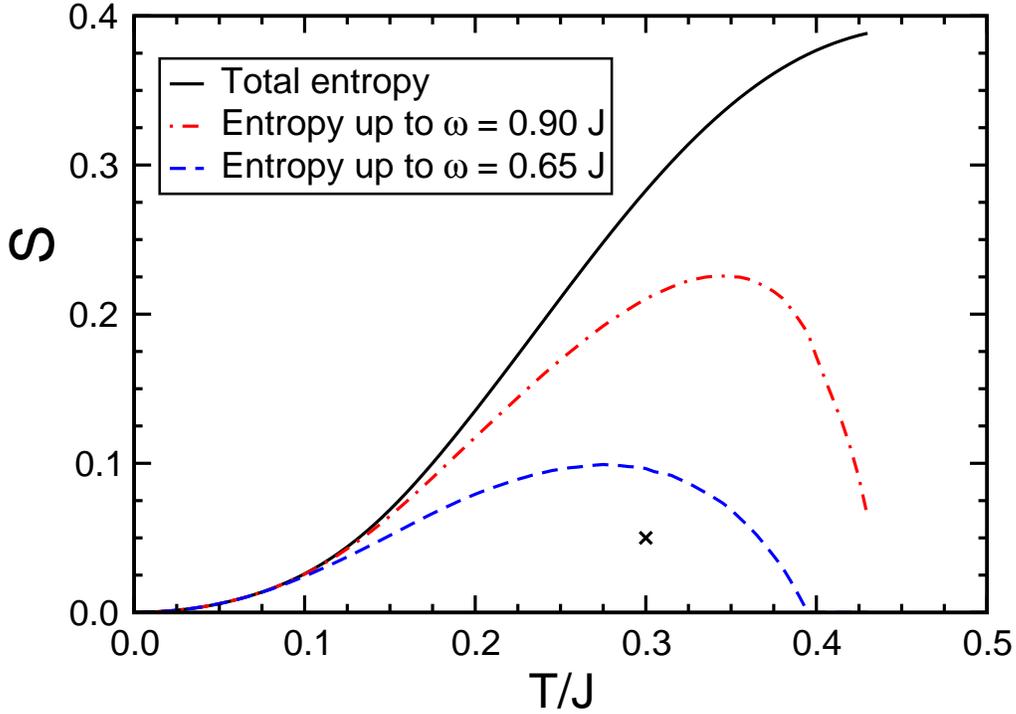}
\caption{Temperature dependence of the RSBMF entropy computed for different range of energies. The cross 
indicates the entropy value predicted by the 
non linear sigma model (NL$\sigma$M) at $T=0.3J$.}
\label{fig4}
\end{center}
\end{figure}

To discern between the contribution of the magnonic low energy excitation and the high energy roton-like excitations 
we have computed the entropy for different ranges of energy \cite{Zheng06}. The blue dashed line of figure \ref{fig4} 
represents the contribution of the low energy modes while the red dotted-dashed line represents the low energy plus 
the roton modes. 
In agreement with \cite{Zheng06} it can be observed that the increasing contribution of the roton modes starts 
from  $T\sim 0.1J$. As a reference we have included the entropy value predicted by the NL$\sigma$M at $T=0.3 J$ (black 
cross) which agrees reasonably with the low energy contribution of the RSBMF. 
The apparently counter-intuitive decreasing of the partial entropies at larger temperatures is due to the temperature 
dependence of $\overline{\omega}_{\bf k}$.      
In the case of a temperature independent dispersion \cite{Zheng06} it is expected an increasing contribution for all 
energy intervals because of the increasing occupation
number average for each mode. In our case, instead, the effect of temperature is to push up an important amount of 
modes outside the corresponding energy intervals, lowering its contribution to the partial entropy for  $T\ge0.3J$. For 
instance, the minimum of the dispersion  at $T=0.4J$ is greater than $0.65J$, then  the contribution of this energy 
interval to the entropy is zero. Similarly, within the energy interval $0-0.9J$, the decreasing of the entropy at $T=0.4J$ 
is due to the fact that the rotonic part of the dispersion flattens and crosses the top of the interval $0-0.9J$.
  
Taking into account the above mentioned we conclude that at least within the range $0.1J-0.3J$ the contribution of the 
rotonic modes becomes 
relevant to the total entropy. In particular, these results confirm the idea that the higher values of entropy found 
with HTE around $T\sim0.3J$ can be attributed to the contribution of the roton excitations. Recently, we have related 
the roton-like excitations with AF collinear fluctuations above the $120^{\circ}$ N\'eel order \cite{Mezio11}. In fact, 
it is easy to check that by including frustrating interactions to second neighbors ($J_2$) the roton excitations soften, 
becoming the precursor of the transition to a collinear state at $J_2/J\sim 0.16$ \cite{Manuel99}.  Then, within the context 
of the RSBMF, the high values of entropy at low temperatures can be related to the contribution of AF collinear fluctuations 
signalled by the flattening of the dispersion relation around the midpoint of the edges of the triangular Brillouin zone. This 
effect is missed in the NL$\sigma$M, where only the effective low energy  modes are taken into account.

\begin{figure}[h]
\begin{center}
\vspace*{0.cm}
\includegraphics*[width=0.6\textwidth,angle=-90]{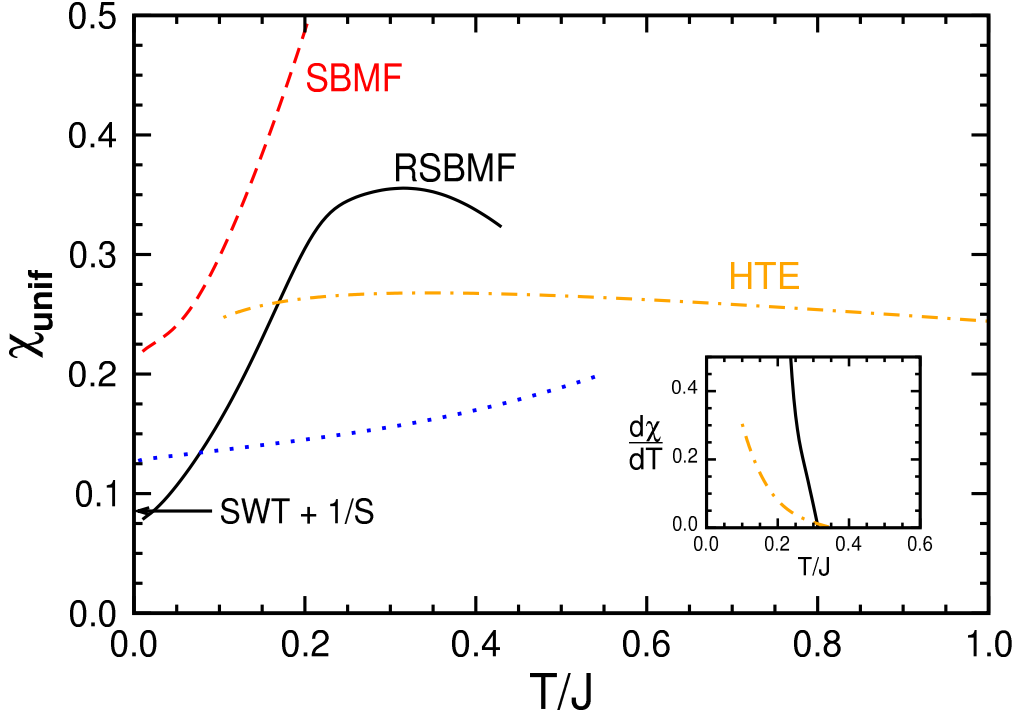}
\caption{Temperature dependence of the uniform susceptibility. RSBMF (\ref{chiRSBMF})(Solid black line); 
SBMF (\ref{chiSBMF}) (red dashed line); HTE of \cite{Eltsner93} (orange dotted dashed line) and SBMF computed 
within the one singlet operator scheme as in \cite{Yoshioka91,Sachdev92} (blue dotted line). The arrow indicates 
the averaged $T=0$ result of the linear spin wave theory plus $1/S$ corrections of \cite{Chubukov94II}. 
Inset: temperature derivative of the 
uniform susceptibility versus temperature.}
\label{fig5}
\end{center}
\end{figure}

The uniform susceptibility calculated within the RSBMF gives
\begin{equation}
\chi_{\rm u}= \frac{S(0)}{T}=\frac{1}{N}\sum_{\bf k} \overline{n}_{\bf k} (\overline{n}_{\bf k}+1)
\label{chiRSBMF}
\end{equation}

\noindent where $S(0)$ is the finite temperature static structure factor 
$S({\bf q})=\sum_{\bf R} <\hat{\bf S}_0.\hat{\bf S}_{\bf R}> \exp(\rmi {\bf q }.{\bf R})$ evaluated at ${\bf q}=0$;   
while in SBMF it results
\begin{equation}
\chi_{\rm u}= \frac{1}{N}\sum_{{\bf k}} n_{{\bf k}} (n_{{\bf k}}+1).
\label{chiSBMF}
\end{equation}
\noindent In figure \ref{fig5} is shown the uniform susceptibility calculated within the RSBMF (black solid line) and 
the SBMF (red dashed line). Consistent
with the previous entropy results, the uniform susceptibility is overestimated by the SBMF. That is, the presence of 
the spurious magnetic excitations enhances the response of the system to a uniform magnetic field. On the other hand, 
the RSBMF results interpolates quite well between the expected zero temperature value (see arrow) and HTE results of 
\cite{Eltsner93} (orange dotted-dashed line). In particular, the extrapolated zero temperature value 
is $\chi_{\rm u}\sim 0.072$ which should be compared with
$\chi_{\rm u}\sim 0.084$, corresponding to the averaged spin wave results plus $1/S$ corrections \cite{Chubukov94}. 
At very low temperatures the RSBMF behaves as $\chi_{\rm u}\sim0.072+0.6 T$ while the behaviour expected for 
classical renormalized regime \cite{Chubukov94} is $\chi_{\rm u}\sim 0.084+0.07 T$. The one order of magnitude 
in the slope is related to the fact that the true correlation length increases faster than the mean field solutions as 
zero temperature is approached (see \cite{Chubukov94,Chubukov95}). For $T>0.15J$ the uniform susceptibility differs from HTE results
 in contrast to the excellent agreement obtained for the entropy. While entropy depends on the correct counting of states, through
the reconstructed dispersion $\overline{\omega}_{\bf k}$, the uniform susceptibility is related to the spectral weight distribution
at ${\bf k}=0$ which at these temperatures, probably, is not correctly taken into account by the mean field approximation.
For instance, we have assumed that the pair of spinons building up the spin-$1$ excitations at ${\bf k}=0$ are sufficiently bound. Therefore, 
the required lowering of the uniform susceptibility to recover the HTE results would be consistent with a weaker bound spinon regime 
for $T>0.15J$. Another source of the above mentioned discrepancy can be attributed to the omission of the continuum contribution to the uniform susceptibility.  
Notice, however, that despite the rounded peak of the HTE is broader than the RSBMF they are both located approximately at the 
same temperature position, $T\sim0.35 J$ \cite{Eltsner93}. This can be noticed in the inset of figure \ref{fig5} where the 
temperature derivative of $\chi_{\rm u}$ versus temperature is plotted. For completeness, we also show the SBMF prediction 
within the one singlet operator scheme (blue dotted line), previously performed in \cite{Yoshioka91,Sachdev92}. In 
this case the interpolation is not good because this scheme of calculation fails to reproduce the low energy spectrum 
of the THM \cite{Mezio11}. 
 \\

Finally, at low temperatures (below $T\!\!\sim\!\!0.1J$), the  specific heat behaves quadratically as $C_v\!\!\sim\!\! a ({T}/{J})^2$, with $a\!\!=\!\! 7.3$ 
and $a\!\!=\!\!5.2$ for SBMF and RSBMF, respectively. As a reference, the latter value should be compared with $a\!=\!5.3 (2)$, 
resulting from an HTE interpolation method \cite{Bernu01}, specially developed for the low temperature specific heat 
behaviour; while $a\!\sim\!3.4$ is found by plugging in the spin wave velocities of the THM in the Debye construction. 

\section{Discussion}
\label{Discussion}
Here we  discuss the validity of the RSBMF developed to compute the low temperature properties of the THM.
Originally, the THM was investigated within SBMF based on the one singlet operator ($A_{ij}$) scheme. As this scheme  gives a 
wrong sum rule, $\int \! \sum_{{\bf k}\alpha}S^{\alpha\alpha}({\bf k},\omega)d\omega= \frac{3}{2}NS(S+1)$, it is well 
known \cite{Arovas88,Auerbach88} that an $\it{ad\;hoc}$ factor of $2/3$ is needed
to compensate the  overcounting of the number of degrees of freedom. For the present two singlet scheme there is no need 
of $\it{ad\;hoc}$ factors since already at the mean field level (SBMF) the sum rule is fulfilled.  Furthermore, it has 
been shown that Gaussian corrections  for the ground state energy and spin stiffness of the THM improve significantly the 
accuracy of the SBMF results when compared with exact diagonalization results on finite systems \cite{Manuel98}.  

\noindent Regarding the magnetic excitations, we have defined a reconstructed dispersion $\overline{\omega}_{\bf k}$ --
guided by the dominant spectral weight of the spin-spin dynamic structure factor-- which reproduces quite well the series 
expansion results. In addition, we found unphysical excitations related to  the fluctuations of the density of Schwinger 
bosons due to the relaxation of the local constraint. Unfortunately, it is difficult to compute the Gaussian corrections 
for the dynamical structure factor  to improve the local constraint. Consequently, to mimic the projection of the SBMF 
into the physical Hilbert space we have discarded the spurious excitations. This allowed us to define the 
reconstructed Schwinger boson mean field, based on bosonic spin-$1$ excitations with the reconstructed 
dispersion $\overline{\omega}_{\bf k}$. 

\noindent Given that the RSBMF thus defined is not rigorously justified       
we have followed an alternative route to confirm whether the counting of the degree of freedom is correctly taken into 
account, at least approximately, within the RSBMF.
To do that we have concentrated on the entropy for the two-spin problem. In this case the problem can be solved exactly and 
in the high temperature limit the entropy is $\mathcal{S}_{\rm ex}=2\ln(2S+1)$. On the other hand, the SBMF prediction can be worked 
out analytically, giving  $\mathcal{S}_{\rm MF}= 2\ln[{(S+1)^{2S+2}}/{S^{2S}}]$. For $S=1/2$, it is found that 
${\mathcal S}_{\rm MF}=2.785\; {\mathcal S}_{\rm ex}$. 
Therefore, the factor $2.785$ gives a faithful compensation of the SBMF entropy due to the wrong counting. Notably, if the SBMF 
entropy (red dashed line) of figure \ref{fig3} is divided by the factor $2.785$ (see black thin solid line of figure \ref{fig3}) 
the RSBMF entropy (black solid line) of figure \ref{fig3} is recovered, at least within the temperature range $0-0.3J$ of interest. 
This agreement  gives further support to our RSBMF procedure.
   
Another possible procedure is to implement the local constraint numerically by using a valence bond basis but sign problems, 
already at the variational level, appear in the Quantum Monte Carlo calculation \cite{Sindzingre94}.
More recently, a  permanent algorithm   was used to compute ground state energies on the kagom\'e lattice within the context of 
projected Schwinger bosons. Although the method seems to circumvent the sign problem, only systems up to $N=36$ size has been 
studied \cite{Tay11}.

\section{Concluding remarks}

We have investigated the low temperature properties of the triangular-lattice Heisenberg model with a bosonic spinon 
theory based on the Schwinger boson mean field theory. Using the two singlet operator scheme of calculation and by 
analyzing the spin-spin and density-density dynamical structure factors we can distinguish between the physical magnetic 
excitations, which reproduce the series expansion results, and the spurious excitations that result from the relaxation of 
the local constraint on the number of bosons per site. By assuming that $\it{i)}$ the effect of projecting into the physical 
Hilbert space can be mimiced by  getting rid of the latter excitations and $\it{ii)}$ that an attractive residual interaction 
binds the spinons,  we obtain a reconstructed Schwinger boson mean field with low energy  spin-$1$  excitations  with 
the reconstructed 'physical' dispersion relation. A comparison with reliable $T=0$ and high temperature ($T\gtrsim 0.3J$) 
results reveal that the RSBMF thus defined provides a very good interpolation for many thermodynamic properties like entropy, 
uniform susceptibility and specific heat over the temperature range $0\leq T\lesssim 0.3J$, which is difficult 
to access by other methods. 
One of our main physical results is to confirm the idea \cite{Zheng06} that the high values of entropy found with 
HTE are due to the excitation of roton excitations which, within the context of our theory, can be identified with the 
collinear short range AF fluctuations above the underlying  $120^{\circ}$ N\'eel correlations \cite{Mezio11}. An alternative interpretation has been proposed 
in \cite{Alicea06}, using the physics of the $XXZ$ Heisenberg model as a starting point. Here a low energy effective theory is constructed in terms 
of fermionized vortices where vortex-antivortex excitations on the honeycomb (dual)
lattice are related to the roton excitations. Interestingly, the dependence of the roton excitations with spatial anisotropy resembles that with 
temperature found in the present work.
Even if the description in terms of fermionized vortices recovers the roton excitations predicted by SE \cite{Zheng06,Zheng06II}, the present  bosonic approach seems to be more appropriate
to describe all the expected features of the isotropic AF Heisenberg model, that is, the $120^{\circ}$ N\'eel order, the correct Goldstone mode structure and the roton
excitations. In any case it would be interesting to investigate the validity range of each approach by  
performing a close comparison between them.
 
The simplicity of the RSBMF along with the consistent description of several features of the THM like  static 
ground state properties, energy spectrum and low temperature thermodynamic properties give strong support
to the bosonic spinon hypothesis to interpret the physics of the 
triangular-lattice Heisenberg model. Of course, the present approximation should be refined by using many body 
diagrammatic techniques \cite{Chubukov91} or by performing $1/{\rm N}$ corrections \cite{Trumper97}. Work in this 
direction is in progress.  

Finally, we believe that a proper extension of the present theory to the anisotropic $XXZ$ case   
would allow one to investigate the unusual magnetic features of 
the recently found \cite{Shirata11,private,Zhou11} inorganic spin-$1/2$ triangular antiferromagnets, $Ba_3CoSb_2O_9$ and $Ba_3CuSb_2O_9$.
An experimental realization of the related $XY$ model has been proposed recently in the area of ultra-cold atoms in optical lattice potentials \cite{Schmied08,Hauke10}.

\ack{This work is supported in part by CONICET (PIP 1948), ANPCyT (PICT R1776) 
and by the US National Science Foundation grant number DMR-1004231.}

\end{document}